\documentclass[prb,twocolumn,showpacs,superscriptaddress]{revtex4}
\usepackage[dvips]{epsfig}
\usepackage{float}
\usepackage{amsmath}
\usepackage{amssymb}
\usepackage{amsfonts}
\usepackage{euscript}
\usepackage{enumerate}
\usepackage{hhline}
\usepackage{threeparttable}
\usepackage{pslatex}
\usepackage{tabularx}
\usepackage{color}
\usepackage{graphicx}% Include figure files
\usepackage{dcolumn}% Align table columns on decimal point
\usepackage{bm}% bold math
\usepackage[sort&compress]{natbib}

\makeatletter
\renewcommand{\@biblabel}[1]{#1. }
\renewcommand{\@dotsep}{500}
\renewcommand{\@pnumwidth}{0em}
\renewcommand{\l@figure}[2]{% #1 is e.g. Figure 1 + caption, #2 is pg.
        \@dottedtocline{1}{1.5em}{2em}{Figure #1}{}\vspace{15pt}}

\begin{document}

%\preprint{Draft 9}

\title{Spectroscopy of 1.55~$\mu$m PbS Quantum Dots on Si Photonic Crystal Cavities with a Fiber Taper Waveguide}% Force line breaks with \\

\author{M. T. Rakher} \email{mrakher@nist.gov}
\affiliation{Center for Nanoscale Science and Technology, National
Institute of Standards and Technology, Gaithersburg, MD 20899, USA}
\author{R. Bose}
\altaffiliation{Current Address: Institute for Research in
Electronics and Applied Physics, University of Maryland, College
Park, MD 20742, USA} \affiliation{Optical Nanostructures Laboratory,
Center for Integrated Science and Engineering, Solid-State Science
and Engineering and Mechanical Engineering, Columbia University, New
York, NY 10027, USA}
\author{C. W. Wong}
\affiliation{Optical Nanostructures Laboratory, Center for
Integrated Science and Engineering, Solid-State Science and
Engineering and Mechanical Engineering, Columbia University, New
York, NY 10027, USA}
\author{K. Srinivasan}
\affiliation{Center for Nanoscale Science and Technology, National
Institute of Standards and Technology, Gaithersburg, MD 20899, USA}

\date{\today}% It is always \today, today,

\begin{abstract}
We use an optical fiber taper waveguide to probe PbS quantum dots
(QDs) dried on Si photonic crystal cavities near 1.55 $\mu$m. We
demonstrate that a low density ($\lesssim 100~\mu$m$^{-2}$) of QDs
does not significantly degrade cavity quality factors as high as
$\approx$3$\times10^4$. We also show that the tapered fiber can be
used to excite the QDs and collect the subsequent cavity-filtered
photoluminescence, and present measurements of reversible
photodarkening and QD saturation. This method represents an
important step towards spectroscopy of single colloidal QDs in the
telecommunications band.
\end{abstract}

 \maketitle

The combination of low optical absorption and mature device
processing has resulted in the development of low loss silicon
photonic devices such as high quality factor ($Q$) photonic crystal
cavities (PCCs) operating in the technologically relevant
1.55~$\mu$m wavelength
range~\cite{ref:Akahane2,ref:Srinivasan7,ref:Song}. Silicon's
indirect bandgap represents a challenge in making light-emitting
devices and as a result there has been considerable interest in
developing hybrid systems integrating a light-emitting
material~\cite{ref:Polman_JAP97,ref:Fang_Bowers_OE05}. Lead salt
colloidal quantum dots \cite{ref:Wise_ACR00,ref:TalapinSci05} (QDs)
represent one such approach.  In addition, their atomic-like
properties suggest the potential for Si-based quantum information
processing in the single QD limit. In this work, we use colloidal
PbS QDs as the active material to interact with Si PCCs with
resonances near 1.55~$\mu$m. Due to the long radiative lifetime
($\approx 700$~ns~\cite{ref:SargentAM05,ref:Rakherunpub}) and small
radiative efficiency of these dried QDs
($\approx1~\%$~\cite{ref:SteckelAM03,ref:SargentAM05}), as well as
challenges associated with measuring low light levels with InGaAs
detectors~\cite{ref:Ribordy}, it is of the utmost importance to
collect as many emitted photons as possible. Previous studies of
PbS/PbSe QDs coupled to Si
microcavities~\cite{ref:Fushman_APL05,ref:BoseAPL07,ref:WuAPL07,ref:PattantyusNANO09}
have relied on free-space micro-photoluminescence methods to pump
and collect the emission from moderately high-$Q$ cavities
($Q\approx10^3$), and have generally operated at relatively high QD
densities, or else have sacrificed spectral resolution to achieve
the count rates needed to operate at a lower QD
density~\cite{ref:BoseAPL09}. In this work, we use an optical fiber
taper waveguide~\cite{ref:Spillane2,ref:Srinivasan7,ref:HwangAPL05}
to couple to the modes of high-$Q$ PCCs ($Q\approx10^4$), thereby
allowing for an efficient out-coupling mechanism for PbS QD
emission. We measure photoluminescence from a low density ($\lesssim
100~\mu$m$^{-2}$) of spun QDs and show that the $Q$ does not degrade
due to QD absorption up to $Q\approx3\times10^4$. We also measure
photodarkening and saturation of the QD emission into the cavity
mode.  This approach may enable the future interrogation of cavity
quantum electrodynamics (cQED) in the PbS/Si system, in much the
same way as has been demonstrated for epitaxial III-V
QDs~\cite{ref:Srinivasan16}.

\begin{figure}[t]
%\centering
        \centerline{\includegraphics[width=8.5cm,clip=true]{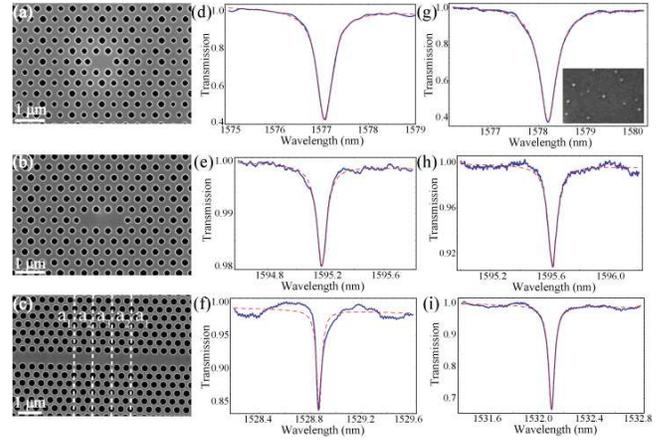}}
    \caption{(a)-(c) SEM images of the H1, L3, and MH cavities, respectively.  The lattice constants in (c) are $\{a_1,a_2,a_3\}$=$\{$410~nm,415~nm,420~nm$\}$.
    (d)-(f) Transmission spectrum of the H1, L3, and MH cavities before QD spin with fits (dashed).
    (d) and (f) were taken with the taper in contact with the cavity, while (e) was taken with the taper above the cavity.
        (g)-(i) Same as (d)-(f), but after QD spin.  Inset to (g): SEM image of QDs in a
        256~nm~$\times$~173~nm area.}%{M.~T. Rakher, et al., APL}
    \label{fig:fig1}
\end{figure}

\begin{figure}
%\centering
        \centerline{\includegraphics[width=8.5cm,
        clip=true]{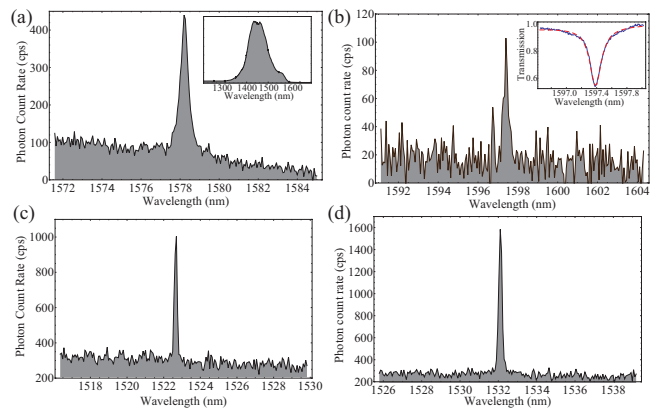}}
    \caption{Fiber-collected PL spectra for (a) H1, (b) L3, and (c)-(d) MH.  Inset to (a): Room-temperature PL of an ensemble of QDs without cavity. Inset to (b): L3 transmission with taper in
contact with cavity.}%{M.~T. Rakher, et al., APL}
    \label{fig:fig2}
\end{figure}

The PbS QDs~\cite{foot} are chemically synthesized
\cite{ref:HinesAdvMat03} and suspended in chloroform. As shown in
the inset of Fig.~\ref{fig:fig2}(a), the emission is centered near
1460~nm with a width of 100~nm due to a combination of size
inhomogeneities and a large homogeneous linewidth at room
temperature. The solution is further diluted with chloroform in a
1:200 mixture. Approximately 20~$\mu$L is spin-coated directly onto
the substrate containing PCCs, yielding an areal density of
$\lesssim 100~\mu$m$^{-2}$ (inset Fig.~\ref{fig:fig1}(g)) as
measured by a scanning electron microscope (SEM). The PCCs measured
(Fig.~\ref{fig:fig1}(a)-(c)) are the well-developed
H1~\cite{ref:Painter3}, L3~\cite{ref:Akahane2}, and
multi-heterostructure (MH) cavities~\cite{ref:Song}, and have been
fabricated in a 250 nm thick Si device layer using standard
silicon-on-insulator fabrication methods. The devices are probed
using an optical fiber taper waveguide, which can be used to measure
the spectral response of the devices in transmission as well to
collect photoluminescence (PL).

Transmission measurements follow the approach of Ref.
\onlinecite{ref:Srinivasan7}, where light from a swept wavelength
external cavity diode laser (1520~nm to 1630~nm) is sent through a
variable optical attenuator and polarization controller before it is
directed through the tapered optical fiber to an InGaAs photodiode.
The taper and sample separation is controlled via $x,y,$ and $z$
stepper stages with 50~nm resolution, and the system is imaged under
a 50X microscope objective. The measurement setup rests in a
$N_2$-rich environment at room temperature to prevent irreversible
photoxidation of the QDs \cite{ref:PetersonPCCP06} and taper
degradation.

This technique enables resonant spectroscopy of the cavity with and
without the active material.  In this way, we measured the cavity
$Q$, before and after addition of the PbS QDs.
Figure~\ref{fig:fig1}(d)-(f) shows a cavity resonance of the H1, L3,
and MH in transmission without QDs. The estimated $Q$ (with
waveguide coupling loss removed\cite{ref:BarclayOE05,ref:Spillane2})
values are 4900, 19~800, and 30~100 respectively.
Figure~\ref{fig:fig1}(g)-(i) shows the cavity's response in
transmission with QDs with corresponding $Q =$ 4500, 23~200, and
29~500.  For these low QD densities, the variation in the extracted
$Q$s due to differences in taper position is greater than the loss
induced by QD absorption, at least up to $Q\approx3\times 10^4$. The
ability to maintain high-$Q$ in the presence of the QDs is promising
for a number of potential applications, such as single QD cQED and
low-threshold microcavity lasers.

For PL measurements, a 980~nm diode laser is coupled through a
variable optical attenuator into the fiber taper, which is brought
into contact with the devices. The transmitted signal is then
directed through a long pass 1064~nm filter and into a grating
spectrometer coupled with a liquid $N_2$ cooled InGaAs array.
Spectra are recorded with a 180~s integration time under a typical
excitation power of 100~$\mu$W. PL spectra from each cavity are
shown in Fig.~\ref{fig:fig2}, including another mode in the MH
cavity that did not appear in transmission (Fig.~\ref{fig:fig2}(c)).
The $Q$ factors observed in PL are consistent with those seen in
transmission measurements, though our spectral resolution is limited
to $\approx$0.09~nm.  We note that the cavity modes operate on the
long wavelength tail-end of the QD distribution, as seen in the
reference PL spectrum shown in the inset of Fig.~\ref{fig:fig2}(a)
for an ensemble of QDs not in a cavity. This suggests the number of
QDs interacting with the cavity modes may be significantly reduced
with respect to the number that physically reside in the cavity,
though a measurement of the QD homogeneous linewidth is needed to
confirm this.

%This suggests that the number of QDs interacting
%with the cavity modes may be significantly reduced with respect to
%the number that physically reside in the cavity, though a
%measurement of the QD homogeneous linewidth is needed to confirm
%this.  It also suggests that QD absorption may play a bigger role in
%reducing cavity $Q$s at shorter wavelengths, though measurements of
%microdisk cavities (not shown) containing a similar density of QDs
%and with resonances near the QD ensemble peak do not show $Q$
%degradation at the level we describe here.

Using the transmission measurements in Fig.~\ref{fig:fig1}, we can
estimate the efficiency $\eta_o$ with which a cavity photon
out-couples into the fiber taper.  A QD's out-coupling efficiency
would then be the product of $\eta_o$ with the fraction of QD
radiation into the cavity mode.  $\eta_o$ is estimated
\cite{ref:BarclayOE05,ref:Spillane2} from the on-resonance
transmission level $T_{res}$ as $\eta_o = (1-\sqrt{T_{res}})/2$ when
the system is in the under-coupled regime and $\eta_{0}$ represents
collection in transmission. For the H1 cavity in
Fig.~\ref{fig:fig1}(g), $T_{res} = 0.381$ so that $\eta_o=19.1~\%$.
A similar efficiency ($T_{res}=0.562$, $\eta_o=12.5~\%$) has been
measured when the taper is in contact with the L3 cavity (inset of
Fig.~\ref{fig:fig2}(b)), while coupling to the MH cavity as shown in
Fig.~\ref{fig:fig1}(i) yields a somewhat smaller value
($T_{res}=0.670$, $\eta_o=9.07~\%$); fluctuations in the detected
signal result in uncertainties in $\eta_o$ of $\leq0.1~\%$. These
results generally compare favorably to calculated free-space
collection efficiencies of $\approx10~\%$ using high numerical
aperture objectives\cite{ref:Tran}, with the added advantage of
direct collection into a single mode optical fiber.

%In general, these results compare very favorably to free-space
%collection with a microscope objective, which for $NA=0.7$ can
%maximally collect $\eta_o=11.1~\%$ of the emission from a randomly
%oriented dipole in vacuum.  The actual measured photon count rate
%will also depend on taper loss and insertion loss, which for these
%cavities was $\approx 5$~dB, though free-space optics approaches
%will also suffer additional loss. Finally, we note that a

\begin{figure}
%\centering
        \centerline{\includegraphics[width=8.5cm,
        clip=true]{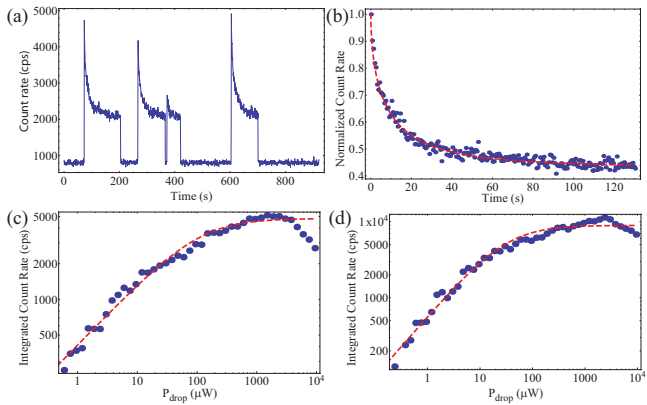}}
    \caption{(a) Continuous measurement of PL on an SPCM while the excitation is intermittently turned off and on. (b)  Close-up of one of the photodarkening curves
    taken in (a) along with fit (dashed). (c), (d) PL saturation measurements of the modes at 1522.7~nm and 1532.1~nm in the MH cavity with fits
    (dashed).}%{M.~T. Rakher, et al., APL}
    \label{fig:fig3}
\end{figure}

Our experimental configuration also enabled measurement of
photodarkening behavior previously observed in PbS QDs
\cite{ref:PetersonPCCP06}.  In this case, PL from the MH cavity is
directed through long pass filters at 1064~nm and 1400~nm and
detected at an InGaAs single photon counting module
(SPCM)\cite{ref:Ribordy} with 2.5~ns gate width, 20~$\%$ detection
efficency, and 5~$\mu$s dead time. As shown in
Fig.~\ref{fig:fig3}(a), the PL is monitored continuously with a
0.6~s integration time while the 980~nm excitation source is turned
on ($P_{drop}=154.0~\mu$W$~\pm9.5~\mu$W) and off. The PL clearly
decays with time and requires an off time of at least 150~s to
completely recover. This kind of photodarkening has been attributed
to an average of single particle blinking where the overall ensemble
PL decreases with time due to increasing numbers of emitters
transitioning to a long-lived dark state
\cite{ref:ChungPRB04,ref:PeltonAPL04,ref:TangJCP05}.
Fig.~\ref{fig:fig3}(b) shows a normalized photodarkening trace taken
under the same excitation conditions as (a).  The data has been fit
with a stretched exponential \cite{ref:TangJCP05}, $I(t) =
I_{eq}+(1-I_{eq}) \textrm{exp}[-(t/T_o)^{\alpha}] $, yielding fit
parameters with 95~$\%$ confidence intervals $I_{eq}=0.435\pm0.006$,
$T_o=8.84\pm0.54$, and $\alpha=0.57\pm0.38$. While the fit
parameters $T_o$ and $\alpha$ are consistent with literature
\cite{ref:TangJCP05}, the actual physical parameters associated with
QD blinking can only be determined with further single QD
measurements beyond the scope of this work. However, $I_{eq}$ is
directly related to the ratio of average time spent in the dark
state to the bright state, which for our QDs computes to a value of
$1.30\pm0.03$. Short-timescale photodarkening and the difficulties
associated with detection at 1.55~$\mu$m make low density QD
measurements that much more challenging.

The final experiment we performed was a saturation spectroscopy
measurement of the two modes of the MH cavity.  In this measurement,
a PL spectrum was recorded (60~s integration) as the dropped
excitation power was increased over more than four decades.  To
avoid photodarkening effects, the excitation was blocked for 30~s
after each measurement and the spectrum was taken only after the
excitation had been on for 30~s.  Two lorentzians were fit to each
spectrum and the integrated count rate under each peak is plotted as
a function of dropped power in Fig.~\ref{fig:fig3}(c),(d). Each of
these curves was fit to a two-level saturation with an adjustable
power dependence, $I(P) = A [P/(P+P_{sat})]^b$. Interestingly, the
saturation curves display a clear sub-linear dependence on the
dropped power below saturation.  The mode at 1522.7~nm (1532.1~nm)
fits to a value of $b=0.518\pm0.046$ ($b=0.795\pm0.082$).  This
sub-linear dependence could be symptomatic of the trapped states
associated with blinking \cite{ref:Babentsov}. The saturation curves
are truncated due to heating in the tapered fiber and in the Si at
excitation powers near 2~mW as evidenced by few nm redshifts of the
cavity modes.  Nonetheless, the saturation power can still be
extracted from the data, albeit with a large uncertainty.  We fit to
$P_{sat} = 153.2~\mu$W$~\pm65.3~\mu$W ($P_{sat} =
42.6~\mu$W$~\pm17.2~\mu$W) for the mode at 1522.7~nm (1532.1~nm).
For a single PbS QD with absorption cross-section
\cite{ref:CademartiriJACS06} $\sigma= 4.59\times10^{-16}$~cm$^2$ and
room-temperature excited state lifetime of $\approx100$~ns
\cite{ref:Rakherunpub}, the expected saturation excitation power for
our tapered fiber setup is $\approx22~\mu$W.  Because the
cross-section is so low, a non-diminished pump approximation is
valid and the single particle saturation power should be accurate
for small QD densities.  Given the uncertainties in the fits as well
as in the values for the cross-section and lifetime, the extracted
saturation powers seem quite reasonable.

In conclusion, we have performed spectroscopy of $1.55~\mu$m PbS QDs
dried on Si photonic crystal cavities using a fiber taper waveguide.
Future experiments will build towards single QD spectroscopy by
lowering the QD density and improving the radiative efficiency by
working in cryogenic conditions\cite{ref:Rakherunpub} and/or using
brighter and more stable colloidal QDs\cite{ref:Pietryga}.  A
combination of these strategies will lead to the development of
novel and useful active nanophotonic devices in the
telecommunications band.

The authors acknowledge fabrication support from D.~L. Kwong and M.
Yu at the Institute of Microelectronics in Singapore, partial
funding support from NSF ECCS 0747787 and the New York State
Foundation for Science, Technology, and Innovation, and useful
discussion with Marcelo Davan\c{c}o at NIST.


\begin{thebibliography}{31}
\expandafter\ifx\csname
natexlab\endcsname\relax\def\natexlab#1{#1}\fi
\expandafter\ifx\csname bibnamefont\endcsname\relax
  \def\bibnamefont#1{#1}\fi
\expandafter\ifx\csname bibfnamefont\endcsname\relax
  \def\bibfnamefont#1{#1}\fi
\expandafter\ifx\csname citenamefont\endcsname\relax
  \def\citenamefont#1{#1}\fi
\expandafter\ifx\csname url\endcsname\relax
  \def\url#1{\texttt{#1}}\fi
\expandafter\ifx\csname urlprefix\endcsname\relax\def\urlprefix{URL
}\fi \providecommand{\bibinfo}[2]{#2}
\providecommand{\eprint}[2][]{\url{#2}}

\bibitem[{\citenamefont{Akahane et~al.}(2003)\citenamefont{Akahane, Asano,
  Song, and Noda}}]{ref:Akahane2}
\bibinfo{author}{\bibfnamefont{Y.}~\bibnamefont{Akahane}},
  \bibinfo{author}{\bibfnamefont{T.}~\bibnamefont{Asano}},
  \bibinfo{author}{\bibfnamefont{B.-S.} \bibnamefont{Song}}, \bibnamefont{and}
  \bibinfo{author}{\bibfnamefont{S.}~\bibnamefont{Noda}},
  \bibinfo{journal}{Nature} \textbf{\bibinfo{volume}{425}},
  \bibinfo{pages}{944} (\bibinfo{year}{2003}).

\bibitem[{\citenamefont{Srinivasan et~al.}(2004)\citenamefont{Srinivasan,
  Barclay, Borselli, and Painter}}]{ref:Srinivasan7}
\bibinfo{author}{\bibfnamefont{K.}~\bibnamefont{Srinivasan}},
  \bibinfo{author}{\bibfnamefont{P.~E.} \bibnamefont{Barclay}},
  \bibinfo{author}{\bibfnamefont{M.}~\bibnamefont{Borselli}}, \bibnamefont{and}
  \bibinfo{author}{\bibfnamefont{O.}~\bibnamefont{Painter}},
  \bibinfo{journal}{Phys. Rev. B} \textbf{\bibinfo{volume}{70}},
  \bibinfo{pages}{081306R} (\bibinfo{year}{2004}).

\bibitem[{\citenamefont{Song et~al.}(2005)\citenamefont{Song, Noda, Asano, and
  Akahane}}]{ref:Song}
\bibinfo{author}{\bibfnamefont{B.-S.} \bibnamefont{Song}},
  \bibinfo{author}{\bibfnamefont{S.}~\bibnamefont{Noda}},
  \bibinfo{author}{\bibfnamefont{T.}~\bibnamefont{Asano}}, \bibnamefont{and}
  \bibinfo{author}{\bibfnamefont{Y.}~\bibnamefont{Akahane}},
  \bibinfo{journal}{Nature Materials} \textbf{\bibinfo{volume}{4}},
  \bibinfo{pages}{207} (\bibinfo{year}{2005}).

\bibitem[{\citenamefont{Polman}(1997)}]{ref:Polman_JAP97}
\bibinfo{author}{\bibfnamefont{A.}~\bibnamefont{Polman}}, \bibinfo{journal}{J.
  Appl. Phys.} \textbf{\bibinfo{volume}{82}}, \bibinfo{pages}{1}
  (\bibinfo{year}{1997}).

\bibitem[{\citenamefont{Park et~al.}(2005)\citenamefont{Park, Fang, Kodama, and
  Bowers}}]{ref:Fang_Bowers_OE05}
\bibinfo{author}{\bibfnamefont{H.}~\bibnamefont{Park}},
  \bibinfo{author}{\bibfnamefont{A.}~\bibnamefont{Fang}},
  \bibinfo{author}{\bibfnamefont{S.}~\bibnamefont{Kodama}}, \bibnamefont{and}
  \bibinfo{author}{\bibfnamefont{J.}~\bibnamefont{Bowers}},
  \bibinfo{journal}{Opt. Express} \textbf{\bibinfo{volume}{13}},
  \bibinfo{pages}{9460} (\bibinfo{year}{2005}).

\bibitem[{\citenamefont{Wise}(2000)}]{ref:Wise_ACR00}
\bibinfo{author}{\bibfnamefont{F.}~\bibnamefont{Wise}}, \bibinfo{journal}{Acc.
  Chem. Res.} \textbf{\bibinfo{volume}{33}}, \bibinfo{pages}{773}
  (\bibinfo{year}{2000}).

\bibitem[{\citenamefont{Talapin and Murray}(2005)}]{ref:TalapinSci05}
\bibinfo{author}{\bibfnamefont{D.~V.} \bibnamefont{Talapin}} \bibnamefont{and}
  \bibinfo{author}{\bibfnamefont{C.~B.} \bibnamefont{Murray}},
  \bibinfo{journal}{Science} \textbf{\bibinfo{volume}{310}},
  \bibinfo{pages}{86} (\bibinfo{year}{2005}).

\bibitem[{\citenamefont{Sargent}(2004)}]{ref:SargentAM05}
\bibinfo{author}{\bibfnamefont{E.~H.} \bibnamefont{Sargent}},
  \bibinfo{journal}{Advanced Materials (Weinheim, Ger.)}
  \textbf{\bibinfo{volume}{17}}, \bibinfo{eid}{515} (\bibinfo{year}{2004}).

\bibitem[{\citenamefont{Rakher et~al.}(2009)\citenamefont{Rakher, Wong, and
  Srinivasan}}]{ref:Rakherunpub}
\bibinfo{author}{\bibfnamefont{M.~T.} \bibnamefont{Rakher}},
  \bibinfo{author}{\bibfnamefont{C.~W.} \bibnamefont{Wong}}, \bibnamefont{and}
  \bibinfo{author}{\bibfnamefont{K.}~\bibnamefont{Srinivasan}},
  \bibinfo{journal}{in preparation}  (\bibinfo{year}{2009}).

\bibitem[{\citenamefont{Steckel et~al.}(2003)\citenamefont{Steckel,
  Coe-Sullivan, Bulovi\'{c}, and Bawendi}}]{ref:SteckelAM03}
\bibinfo{author}{\bibfnamefont{J.~S.} \bibnamefont{Steckel}},
  \bibinfo{author}{\bibfnamefont{S.}~\bibnamefont{Coe-Sullivan}},
  \bibinfo{author}{\bibfnamefont{V.}~\bibnamefont{Bulovi\'{c}}},
  \bibnamefont{and} \bibinfo{author}{\bibfnamefont{M.~G.}
  \bibnamefont{Bawendi}}, \bibinfo{journal}{Advanced Materials (Weinheim,
  Ger.)} \textbf{\bibinfo{volume}{15}}, \bibinfo{eid}{1862}
  (\bibinfo{year}{2003}).

\bibitem[{\citenamefont{Ribordy et~al.}(2004)\citenamefont{Ribordy, Gisin,
  Guinnard, Stucki, Wegmuller, and Zbinden}}]{ref:Ribordy}
\bibinfo{author}{\bibfnamefont{G.}~\bibnamefont{Ribordy}},
  \bibinfo{author}{\bibfnamefont{N.}~\bibnamefont{Gisin}},
  \bibinfo{author}{\bibfnamefont{O.}~\bibnamefont{Guinnard}},
  \bibinfo{author}{\bibfnamefont{D.}~\bibnamefont{Stucki}},
  \bibinfo{author}{\bibfnamefont{M.}~\bibnamefont{Wegmuller}},
  \bibnamefont{and} \bibinfo{author}{\bibfnamefont{H.}~\bibnamefont{Zbinden}},
  \bibinfo{journal}{Journal of Modern Optics} \textbf{\bibinfo{volume}{51}},
  \bibinfo{pages}{1381} (\bibinfo{year}{2004}).

\bibitem[{\citenamefont{Fushman et~al.}(2005)\citenamefont{Fushman, Englund,
  and Vuckovic}}]{ref:Fushman_APL05}
\bibinfo{author}{\bibfnamefont{I.}~\bibnamefont{Fushman}},
  \bibinfo{author}{\bibfnamefont{D.}~\bibnamefont{Englund}}, \bibnamefont{and}
  \bibinfo{author}{\bibfnamefont{J.}~\bibnamefont{Vuckovic}},
  \bibinfo{journal}{Appl. Phys. Lett.} \textbf{\bibinfo{volume}{87}}
  (\bibinfo{year}{2005}).

\bibitem[{\citenamefont{Bose et~al.}(2007)\citenamefont{Bose, Yang, Chatterjee,
  Gao, and Wong}}]{ref:BoseAPL07}
\bibinfo{author}{\bibfnamefont{R.}~\bibnamefont{Bose}},
  \bibinfo{author}{\bibfnamefont{X.}~\bibnamefont{Yang}},
  \bibinfo{author}{\bibfnamefont{R.}~\bibnamefont{Chatterjee}},
  \bibinfo{author}{\bibfnamefont{J.}~\bibnamefont{Gao}}, \bibnamefont{and}
  \bibinfo{author}{\bibfnamefont{C.~W.} \bibnamefont{Wong}},
  \bibinfo{journal}{Applied Physics Letters} \textbf{\bibinfo{volume}{90}},
  \bibinfo{eid}{111117} (\bibinfo{year}{2007}).

\bibitem[{\citenamefont{Wu et~al.}(2007)\citenamefont{Wu, Mi, Bhattacharya,
  Zhu, and Xu}}]{ref:WuAPL07}
\bibinfo{author}{\bibfnamefont{Z.}~\bibnamefont{Wu}},
  \bibinfo{author}{\bibfnamefont{Z.}~\bibnamefont{Mi}},
  \bibinfo{author}{\bibfnamefont{P.}~\bibnamefont{Bhattacharya}},
  \bibinfo{author}{\bibfnamefont{T.}~\bibnamefont{Zhu}}, \bibnamefont{and}
  \bibinfo{author}{\bibfnamefont{J.}~\bibnamefont{Xu}},
  \bibinfo{journal}{Applied Physics Letters} \textbf{\bibinfo{volume}{90}},
  \bibinfo{eid}{171105} (\bibinfo{year}{2007}).

\bibitem[{\citenamefont{Pattantyus-Abraham
  et~al.}(2009)\citenamefont{Pattantyus-Abraham, Qiao, Shan, Abel, Wang, van
  Veggel, and Young}}]{ref:PattantyusNANO09}
\bibinfo{author}{\bibfnamefont{A.~G.} \bibnamefont{Pattantyus-Abraham}},
  \bibinfo{author}{\bibfnamefont{H.}~\bibnamefont{Qiao}},
  \bibinfo{author}{\bibfnamefont{J.}~\bibnamefont{Shan}},
  \bibinfo{author}{\bibfnamefont{K.~A.} \bibnamefont{Abel}},
  \bibinfo{author}{\bibfnamefont{T.-S.} \bibnamefont{Wang}},
  \bibinfo{author}{\bibfnamefont{F.~C. J.~M.} \bibnamefont{van Veggel}},
  \bibnamefont{and} \bibinfo{author}{\bibfnamefont{J.~F.} \bibnamefont{Young}},
  \bibinfo{journal}{Nano Letters} \textbf{\bibinfo{volume}{9}},
  \bibinfo{eid}{2849} (\bibinfo{year}{2009}).

\bibitem[{\citenamefont{Bose et~al.}(2009)\citenamefont{Bose, McMillan, Gao,
  and Wong}}]{ref:BoseAPL09}
\bibinfo{author}{\bibfnamefont{R.}~\bibnamefont{Bose}},
  \bibinfo{author}{\bibfnamefont{J.~F.} \bibnamefont{McMillan}},
  \bibinfo{author}{\bibfnamefont{J.}~\bibnamefont{Gao}}, \bibnamefont{and}
  \bibinfo{author}{\bibfnamefont{C.~W.} \bibnamefont{Wong}},
  \bibinfo{journal}{Applied Physics Letters} \textbf{\bibinfo{volume}{95}},
  \bibinfo{eid}{131112} (\bibinfo{year}{2009}).

\bibitem[{\citenamefont{Spillane et~al.}(2003)\citenamefont{Spillane,
  Kippenberg, Painter, and Vahala}}]{ref:Spillane2}
\bibinfo{author}{\bibfnamefont{S.~M.} \bibnamefont{Spillane}},
  \bibinfo{author}{\bibfnamefont{T.~J.} \bibnamefont{Kippenberg}},
  \bibinfo{author}{\bibfnamefont{O.~J.} \bibnamefont{Painter}},
  \bibnamefont{and} \bibinfo{author}{\bibfnamefont{K.~J.}
  \bibnamefont{Vahala}}, \bibinfo{journal}{Phys. Rev. Lett.}
  \textbf{\bibinfo{volume}{91}}, \bibinfo{pages}{043902}
  (\bibinfo{year}{2003}).

\bibitem[{\citenamefont{Hwang et~al.}(2005)\citenamefont{Hwang, Kim, Yang, Kim,
  Lee, and Lee}}]{ref:HwangAPL05}
\bibinfo{author}{\bibfnamefont{I.-K.} \bibnamefont{Hwang}},
  \bibinfo{author}{\bibfnamefont{S.-K.} \bibnamefont{Kim}},
  \bibinfo{author}{\bibfnamefont{J.-K.} \bibnamefont{Yang}},
  \bibinfo{author}{\bibfnamefont{S.-H.} \bibnamefont{Kim}},
  \bibinfo{author}{\bibfnamefont{S.~H.} \bibnamefont{Lee}}, \bibnamefont{and}
  \bibinfo{author}{\bibfnamefont{Y.-H.} \bibnamefont{Lee}},
  \bibinfo{journal}{Appl. Phys. Lett.} \textbf{\bibinfo{volume}{87}},
  \bibinfo{eid}{131107} (\bibinfo{year}{2005}).

\bibitem[{\citenamefont{Srinivasan and Painter}(2007)}]{ref:Srinivasan16}
\bibinfo{author}{\bibfnamefont{K.}~\bibnamefont{Srinivasan}} \bibnamefont{and}
  \bibinfo{author}{\bibfnamefont{O.}~\bibnamefont{Painter}},
  \bibinfo{journal}{Nature (London)} \textbf{\bibinfo{volume}{450}},
  \bibinfo{pages}{862} (\bibinfo{year}{2007}).

\bibitem[{\citenamefont{Srinivasan and Painter}(2007)}]{foot}
\bibinfo{author}{Purchased from Evident Technologies and identified in this paper to
foster understanding, without implying recommendation or endorsement
by NIST}.

\bibitem[{\citenamefont{Hines and Scholes}(2003)}]{ref:HinesAdvMat03}
\bibinfo{author}{\bibfnamefont{M.~A.} \bibnamefont{Hines}} \bibnamefont{and}
  \bibinfo{author}{\bibfnamefont{G.~D.} \bibnamefont{Scholes}},
  \bibinfo{journal}{Adv. Mater. (Weinheim, Ger.)}
  \textbf{\bibinfo{volume}{15}}, \bibinfo{eid}{1844} (\bibinfo{year}{2003}).

\bibitem[{\citenamefont{Painter et~al.}(1999)\citenamefont{Painter, Lee, Yariv,
  Scherer, O'Brien, Dapkus, and Kim}}]{ref:Painter3}
\bibinfo{author}{\bibfnamefont{O.}~\bibnamefont{Painter}},
  \bibinfo{author}{\bibfnamefont{R.~K.} \bibnamefont{Lee}},
  \bibinfo{author}{\bibfnamefont{A.}~\bibnamefont{Yariv}},
  \bibinfo{author}{\bibfnamefont{A.}~\bibnamefont{Scherer}},
  \bibinfo{author}{\bibfnamefont{J.~D.} \bibnamefont{O'Brien}},
  \bibinfo{author}{\bibfnamefont{P.~D.} \bibnamefont{Dapkus}},
  \bibnamefont{and} \bibinfo{author}{\bibfnamefont{I.}~\bibnamefont{Kim}},
  \bibinfo{journal}{Science} \textbf{\bibinfo{volume}{284}},
  \bibinfo{pages}{1819} (\bibinfo{year}{1999}).

\bibitem[{\citenamefont{Peterson and Krauss}(2006)}]{ref:PetersonPCCP06}
\bibinfo{author}{\bibfnamefont{J.~J.} \bibnamefont{Peterson}} \bibnamefont{and}
  \bibinfo{author}{\bibfnamefont{T.~D.} \bibnamefont{Krauss}},
  \bibinfo{journal}{Phys. Chem. Chem. Phys.} \textbf{\bibinfo{volume}{8}},
  \bibinfo{eid}{3851} (\bibinfo{year}{2006}).

\bibitem[{\citenamefont{Barclay et~al.}(2005)\citenamefont{Barclay, Srinivasan,
  and Painter}}]{ref:BarclayOE05}
\bibinfo{author}{\bibfnamefont{P.}~\bibnamefont{Barclay}},
  \bibinfo{author}{\bibfnamefont{K.}~\bibnamefont{Srinivasan}},
  \bibnamefont{and} \bibinfo{author}{\bibfnamefont{O.}~\bibnamefont{Painter}},
  \bibinfo{journal}{Opt. Express} \textbf{\bibinfo{volume}{13}},
  \bibinfo{pages}{801} (\bibinfo{year}{2005}).

\bibitem[{\citenamefont{Tran et~al.}(2009)\citenamefont{Tran, Combri\'{e}, and
  Rossi}}]{ref:Tran}
\bibinfo{author}{\bibfnamefont{N.-V.-Q.} \bibnamefont{Tran}},
  \bibinfo{author}{\bibfnamefont{S.}~\bibnamefont{Combri\'{e}}},
  \bibnamefont{and} \bibinfo{author}{\bibfnamefont{A.~D.} \bibnamefont{Rossi}},
  \bibinfo{journal}{Phys. Rev. B} \textbf{\bibinfo{volume}{79}},
  \bibinfo{eid}{041101} (\bibinfo{year}{2009}).

\bibitem[{\citenamefont{Chung and Bawendi}(2004)}]{ref:ChungPRB04}
\bibinfo{author}{\bibfnamefont{I.}~\bibnamefont{Chung}} \bibnamefont{and}
  \bibinfo{author}{\bibfnamefont{M.~G.} \bibnamefont{Bawendi}},
  \bibinfo{journal}{Phys. Rev. B} \textbf{\bibinfo{volume}{70}},
  \bibinfo{pages}{165304} (\bibinfo{year}{2004}).

\bibitem[{\citenamefont{Pelton et~al.}(2004)\citenamefont{Pelton, Grier, and
  Guyot-Sionnest}}]{ref:PeltonAPL04}
\bibinfo{author}{\bibfnamefont{M.}~\bibnamefont{Pelton}},
  \bibinfo{author}{\bibfnamefont{D.~G.} \bibnamefont{Grier}}, \bibnamefont{and}
  \bibinfo{author}{\bibfnamefont{P.}~\bibnamefont{Guyot-Sionnest}},
  \bibinfo{journal}{Applied Physics Letters} \textbf{\bibinfo{volume}{85}},
  \bibinfo{pages}{819} (\bibinfo{year}{2004}).

\bibitem[{\citenamefont{Tang and Marcus}(2005)}]{ref:TangJCP05}
\bibinfo{author}{\bibfnamefont{J.}~\bibnamefont{Tang}} \bibnamefont{and}
  \bibinfo{author}{\bibfnamefont{R.~A.} \bibnamefont{Marcus}},
  \bibinfo{journal}{The Journal of Chemical Physics}
  \textbf{\bibinfo{volume}{123}}, \bibinfo{eid}{204511} (\bibinfo{year}{2005}).

\bibitem[{\citenamefont{Babentsov et~al.}(2005)\citenamefont{Babentsov,
  Riegler, Schneider, Fiederle, and Nann}}]{ref:Babentsov}
\bibinfo{author}{\bibfnamefont{V.}~\bibnamefont{Babentsov}},
  \bibinfo{author}{\bibfnamefont{J.}~\bibnamefont{Riegler}},
  \bibinfo{author}{\bibfnamefont{J.}~\bibnamefont{Schneider}},
  \bibinfo{author}{\bibfnamefont{M.}~\bibnamefont{Fiederle}}, \bibnamefont{and}
  \bibinfo{author}{\bibfnamefont{T.}~\bibnamefont{Nann}}, \bibinfo{journal}{J.
  Phys. Chem. B} \textbf{\bibinfo{volume}{109}}, \bibinfo{pages}{15349}
  (\bibinfo{year}{2005}).

\bibitem[{\citenamefont{Cademartiri et~al.}(2006)\citenamefont{Cademartiri,
  Montanari, Calestani, Migliori, Guagliardi, and
  Ozin}}]{ref:CademartiriJACS06}
\bibinfo{author}{\bibfnamefont{L.}~\bibnamefont{Cademartiri}},
  \bibinfo{author}{\bibfnamefont{E.}~\bibnamefont{Montanari}},
  \bibinfo{author}{\bibfnamefont{G.}~\bibnamefont{Calestani}},
  \bibinfo{author}{\bibfnamefont{A.}~\bibnamefont{Migliori}},
  \bibinfo{author}{\bibfnamefont{A.}~\bibnamefont{Guagliardi}},
  \bibnamefont{and} \bibinfo{author}{\bibfnamefont{G.~A.} \bibnamefont{Ozin}},
  \bibinfo{journal}{Journal of the American Chemical Society}
  \textbf{\bibinfo{volume}{128}}, \bibinfo{eid}{10337} (\bibinfo{year}{2006}).

\bibitem[{\citenamefont{Pietryga et~al.}(2008)\citenamefont{Pietryga, Werder,
  Williams, Casson, Schaller, Klimov, and Hollingsworth}}]{ref:Pietryga}
\bibinfo{author}{\bibfnamefont{J.~M.} \bibnamefont{Pietryga}},
  \bibinfo{author}{\bibfnamefont{D.~J.} \bibnamefont{Werder}},
  \bibinfo{author}{\bibfnamefont{D.~J.} \bibnamefont{Williams}},
  \bibinfo{author}{\bibfnamefont{J.~L.} \bibnamefont{Casson}},
  \bibinfo{author}{\bibfnamefont{R.~D.} \bibnamefont{Schaller}},
  \bibinfo{author}{\bibfnamefont{V.~I.} \bibnamefont{Klimov}},
  \bibnamefont{and} \bibinfo{author}{\bibfnamefont{J.~A.}
  \bibnamefont{Hollingsworth}}, \bibinfo{journal}{J. Am. Chem. Soc.}
  \textbf{\bibinfo{volume}{130}}, \bibinfo{eid}{4879} (\bibinfo{year}{2008}).

\end{thebibliography}
\end{document}